\documentclass[%
 aip,
 amsmath,amssymb,
 reprint,%
]{revtex4-1}

\usepackage{graphicx}
\usepackage{dcolumn}
\usepackage{bm}

\usepackage[utf8]{inputenc}
\usepackage[T1]{fontenc}
\usepackage{mathptmx}
\usepackage{etoolbox}
\usepackage{xcolor}
\usepackage{hyperref}

\makeatletter
\def\@email#1#2{%
 \endgroup
 \patchcmd{\titleblock@produce}
  {\frontmatter@RRAPformat}
  {\frontmatter@RRAPformat{\produce@RRAP{*#1\href{mailto:#2}{#2}}}\frontmatter@RRAPformat}
  {}{}
}%
\makeatother
\begin{document}

\preprint{AIP/123-QED}

\title{Machine-Learned Many-Body Potentials for Charged Colloids reveal Gas-Liquid Spinodal Instabilities only in the strong-coupling regime of Primitive Models}

\author{Thijs ter Rele}
\affiliation{Soft Condensed Matter \& Biophysics, Debye Institute for Nanomaterials Science, Utrecht University, Princetonplein 1, 3584 CC Utrecht, The Netherlands}
\email{t.r.terrele@uu.nl}
\author{Ren\'{e} van Roij}
\affiliation{Institute for Theoretical Physics, Utrecht University, Princetonplein 5, 3584 CC Utrecht, The Netherlands}
\author{Marjolein Dijkstra}
\affiliation{Soft Condensed Matter \& Biophysics, Debye Institute for Nanomaterials Science, Utrecht University, Princetonplein 1, 3584 CC Utrecht, The Netherlands}

\date{\today}

\begin{abstract}
Past experimental observations of gas-liquid and gas-crystal coexistence in low-salinity suspensions of highly charged colloids have suggested the existence of like-charge attraction. Evidence for this phenomenon was also observed in  primitive-model simulations of (asymmetric) electrolytes and of low-charge nanoparticle dispersions. These results from low-valency simulations have often been extrapolated to experimental parameter regimes of high colloid valency where like-charge attraction between colloids has been reported. However, direct simulations of highly charged colloids remain computationally demanding. To circumvent slow equilibration, we employ a machine-learning (ML) framework to construct ML potentials that accurately describe the effective colloid interactions. Our ML potentials enable fast simulations of dispersions and successfully reproduce the gas-liquid and gas-solid phase separation observed in primitive-model simulations at low charge numbers. 
Extending the ML-based simulations to higher valencies, where primitive-model simulations become prohibitively slow, also reveals  like-charge attractions and gas-liquid spinodal instabilities, however only in the regime of strongly coupled electrostatic interactions and not in the weakly coupled Poisson-Boltzmann regime of the experimental observations of colloidal like-charge attractions.

\end{abstract}
\maketitle

\section{Introduction}

Charged colloidal systems play a crucial role in applications ranging from paint formulation and medicine development to food science. Given their broad relevance, there has long been significant interest in understanding the structure, phase behavior, and interparticle interactions of these charged colloidal suspensions. Despite many advances over the past few decades, our theoretical understanding remains largely based on  classical models developed in the 1940s. These early works combined principles of electrostatics (via the Poisson equation) with statistical mechanics (via the Boltzmann distribution) to form the Poisson-Boltzmann framework. This framework   underlies  the Derjaguin-Landau-Verwey-Overbeek (DLVO) theory, which describes interactions between charged surfaces in solution.\cite{Derjaguin1941, Verwey1948} 

In index-matched suspensions, where Van der Waals forces from dipole-dipole fluctuations can be neglected, DLVO theory predicts a screened-Coulomb repulsion between pairs of like-charged colloidal particles. Within this mean-field framework, screening of colloidal surface charges is mediated by mobile ions in the background electrolyte, which form an electric double layer (EDL) around each particle. The EDL consists of a local excess of counterions and a corresponding depletion of co-ions near the charged surface, with a characteristic thickness on the order of the Debye length, typically  1-100  nm in aqueous systems, depending on the salt concentration. When the diffuse layers of two colloids overlap at surface separations comparable to several Debye lengths,  a (Yukawa-like) screened-Coulomb repulsion emerges that  is (much) weaker than the bare Coulomb interaction. 

The pairwise DLVO potential, which is an effective coarse-grained ``colloids-only'' description in which the ion degrees of freedom are integrated out, has become a cornerstone of colloid science and has proven highly effective in describing interactions and phase behavior across a wide range of systems.\cite{Levin2002, Israelachvili2011, Trefalt2017} However, since the 1990s, a number of experiments has cast some doubts on the universal validity of DLVO theory, particularly for systems consisting of highly charged, micron-scale particles at low-salt concentrations.\cite{Tata1992,Kepler1994,Wang2024,Ito1994,Tata1997,Royall2003,Tata2006,Larsen1996,Larsen1997} In this low-salt regime, where the Debye length can  exceed ~100 nm, (i) many-body interactions (beyond simple pairwise) are expected to arise due to multiple overlaps of EDLs, and (ii) the effective ionic strength (and thus the effective Debye length) is no longer predominately determined by the background electrolyte but instead influenced by the increasing counterion concentration as the  colloid concentration increases. 

Experimental studies on highly charged colloids under low-salinity conditions revealed unexpected attractive interactions between like-charged particles, in contradiction with DLVO theory. 
Evidence for these anomalous attractions includes observations of  gas-liquid phase coexistence in colloidal suspensions,\cite{Tata1992} clustering of colloidal particles in bulk,\cite{Kepler1994, Wang2024} the formation of voids,\cite{Ito1994, Tata1997, Royall2003, Tata2006} and the unexpected stability of long-lived colloidal crystals.\cite{Larsen1996, Larsen1997} These findings generated intense debate within the colloid community, as several studies questioned the reproducibility of the reported phenomena\cite{Palberg1994_tata} or attributed them to possible experimental artifacts.\cite{Kepler1994, Squires2000_brenner} Despite extensive efforts, no definitive explanation for the observed like-charge attraction has yet been established, although numerous theoretical mechanisms have been proposed.

For colloids with high surface charge densities, like-charge attraction can emerge from electrostatic interactions in the strong-coupling regime, which is characterized by low temperature, low dielectric constant, and high ionic valency.\cite{Kjellander1988, Allahyarov1998, Netz2001, Naji2005, Punkkinen2008, Samaj2011, Palaia2022}  
In this strong-coupling regime,  (pointlike) counterions no longer form a diffuse three-dimensional cloud as predicted by Poisson-Boltzmann theory, but instead condense strongly  onto the highly charged colloidal surfaces, forming a quasi-two-dimensional correlated layer. When two such ion-decorated surfaces approach each other,  correlations among the adsorbed counterions generate an effective attraction between  like-charged colloids.  
For multivalent ions, this type of strong-coupling behavior occurs at higher temperatures than for monovalent ions, since the  strength of ion-ion interactions scales with the square of the ion valency.\cite{Punkkinen2008}

A second class of theories, based on Poisson-Boltzmann mean-field calculations,  attributes the observed like-charge attractions to the cohesive free energy associated with each colloid's own EDL. While this is an irrelevant constant free-energy contribution per colloidal particle in the high-salt regime---where the salt concentration and thus the Debye length are determined by a fixed background electrolyte---it becomes nonlinearly dependent on the colloid packing fraction at low salinity, where the salt concentration (and hence the effective Debye length) is affected by the colloid density. In this low-salt regime, the single-colloid self-energy acquires a cohesive, many-body character that lowers both the osmotic pressure and the colloid chemical potential. For highly charged colloids at low salinity, this cohesive free-energy contribution, encoded in  so-called volume terms, has been predicted to be significant enough to drive phase coexistence between dilute, gas-like  colloidal states and much denser colloidal liquid-like and crystalline states.\cite{vanRoij1997, vanRoij1999, Zoetekouw2006, Warren2000, Russ2002} 
These volume terms have also been observed to provide a stabilizing force for phase separation in suspensions of binary mixtures of like-charged colloids.\cite{Yoshizawa2012} 

More recent experiments from the 2010s have reported clustering of like-charged colloidal particles that, counterintuitively, depends strongly  on both the sign of the colloidal charge and the solvent type.\cite{Gomez2009, Kubincov2020, Wang2024, Wang2025} These findings point to a crucial role of the molecular properties of the solvent, in particular its polarizability and polarization near the colloidal  surface. This prominent role of the solvent is in sharp contrast to the Primitive Model, in which the solvent is treated as a structureless dielectric continuum that simply modulates the Coulomb interaction, and on which both strong-coupling theories and Poisson-Boltzmann theories are usually based. Although   explicit-solvent models are fascinating and deserve further scrutiny,  the present study focuses on primitive models. 

To elucidate the mechanism behind like-charge attractions in primitive models, researchers have  performed fine-grained simulations of charged colloids suspended in  electrolytes. Unlike DLVO theory, which integrates out the ion degrees of freedom, these primitive-model (PM) simulations treat ions explicitly  while representing the solvent as a uniform dielectric continuum. Remarkably, these PM simulations have  reproduced some of the anomalous like-charge attraction phenomena observed experimentally---features that were absent in simulations based solely on pairwise DLVO interactions. For example, Monte Carlo (MC) simulations of asymmetric electrolytes using the Primitive Model, containing only colloids and charge-neutralizing counterions, have revealed clustering of charged particles and gas-liquid phase separation.\cite{Linse1999, Rescic2001} Notably,  gas-liquid phase coexistence emerges only at small charge asymmetries, while gas-crystal phase coexistence persists at larger charge asymmetries.\cite{Hynninen2007, Hynninen2009} Simulations of charged nanoparticles at low temperatures have further demonstrated short-range attractions that deviate from the predictions of the DLVO potential. These simulations involved two isolated colloids suspended in either a 2:2 or 2:1 electrolyte.\cite{Wu1998, Wu1999, Lin2021} However, when colloids were suspended in a 1:1 electrolyte, the interactions closely followed the DLVO potential, suggesting that correlated ion fluctuations in the strong-coupling regime---prominent at higher ion valencies---are responsible for the observed attractions. Interestingly, the phase separations found in simulations of asymmetric, low-charge electrolytes, located in the strongly coupled electrostatic regime, were later extrapolated within the Poisson-Boltzmann framework to explain the experimentally observed phase separation of highly charged colloids.\cite{Zoetekouw2006prl}

Unfortunately, the aforementioned simulation studies were constrained by prohibitively long computation times,\cite{Wu1999, Wu1998, Hynninen2007, Hynninen2009, Lin2021} and as a result, were performed at lower colloid charge numbers than those typically used in experiments. 
Simulations at high charge numbers pose a particular challenge, as they require a significantly larger number of charge-neutralizing counterions. This large particle number, combined with the long-ranged nature of Coulombic interactions, severely slows down primitive-model simulations. For  similar reasons, simulations at high salt concentrations were also avoided. 
However, to rigorously test the hypothesis of volume-term-induced gas-liquid or gas-crystal phase separation, it is essential to perform simulations at the high colloid charge numbers encountered in experimental systems.

In a previous paper, we have presented a Machine-Learning (ML) framework for  generating colloids-only potentials  that accurately reproduce the effective  interactions between colloidal particles in primitive-model simulations.\cite{terRele2025} 
This approach builds on ML techniques that have been successfully applied to speed up atomistic and coarse-grained simulations in earlier studies.\cite{Deringer2019, CamposVillalobos2021, Nguyen2022, Giunta2023, Argun2024, campos2024machine} 
Our previous study focused  on reproducing the interactions between particles in experimental colloidal systems dispersed in low-polar solvents and the effective interactions between ions in an electrolyte.\cite{Royall2006, terRele2025} In this work, we extend this  framework to generate potentials for highly charged colloids suspended in an aqueous 1:1 electrolyte. In Section \ref{sec:assym100} we examine whether these ML potentials can accurately capture the behavior of colloids  in the strong-coupling regime. In Section \ref{sec:assymhigh}, we investigate higher colloid charge valencies and use the ML potentials to study the large-scale phase behavior of colloidal suspensions, focusing in particular on the temperature at which phase separation occurs. Finally, in Section \ref{sec:electrolyte_salt}, we analyze how varying the salt concentration affects  the resulting interaction potentials.

\section{Model and Methods}
\subsection{Primitive-Model simulations}\label{sec:model-method}
We study suspensions of charged colloidal particles that we model by the three-dimensional Primitive Model (PM). The PM consists of larger and smaller charged spheres representing the colloidal particles and the co- and counterions, respectively, while the solvent is treated as a structureless dielectric continuum at temperature $T$.  The potential between the particles is pairwise and consists of a sum of pseudo-hard-sphere repulsions and electrostatic Coulomb interactions. The short-ranged repulsions are described by the Weeks-Chandler-Andersen (WCA) potential, which for two particles $k$ and $l$ at distance $r_{kl}$ is given by
\begin{equation}
    \hspace{-2mm} U_{WCA}(r_{kl}) =\begin{cases}4\overline{\varepsilon}\left[\left(\frac{\sigma_{kl}}{r_{kl}}\right)^{12} \hspace{-2mm} - \left(\frac{\sigma_{kl}}{r_{kl}}\right)^{6} \right] + \overline{\varepsilon},& \  \hspace{-2mm} r_{kl} \leq 2^{1/6} \sigma _{kl};
        \\ 0, & \  \hspace{-2mm} r_{kl} > 2^{1/6} \sigma_{kl}. 
    \end{cases}
\end{equation}
Here $\sigma_{kl}$ is the (quasi) contact distance between two particles and $\beta \overline{\varepsilon} =40$ sets the interaction strength in units of $k_BT$, with $\beta = 1/k_B T$ and $k_B$ the Boltzmann constant. The Coulomb potential between a pair of particles is given by
\begin{equation}
    U_{C}(r_{kl}) = k_B T \frac{Q_k Q_l \lambda_B}{r_{kl}} ,
\end{equation}
where $Q_k$ and $Q_l$ denote the charge valencies of particle $k$ and $l$, respectively, and where $\lambda_B = \beta q^2/(4\pi\epsilon)$ is the Bjerrum length, with $q$ the elementary charge and $\epsilon$ the dielectric permittivity of the solvent. 

\begin{figure}[ht]
  \centering
    \includegraphics[width=\linewidth]{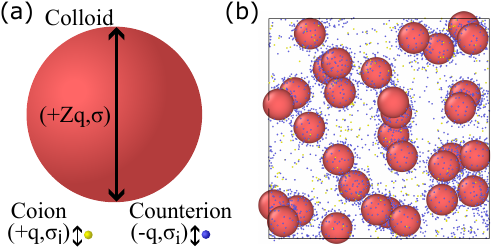}
  \centering
  \caption{(a) Illustration of the three species in the Primitive Model, representing colloids, counterions, and coions with charges $Zq$, $-q$, and $+q$,  and diameters $\sigma$, $\sigma_i$, and  $\sigma_i$, respectively, where $Z>0$ is the colloid valency, $q$ the elementary charge, and $\sigma_i=\sigma/20$ the diameter of the co- and counter-ions.  (b) Configuration of a simulation of $N = 32$ colloids and 3854 ions, used in generating ML potentials. This configuration is representative of a system in the strong-coupling regime. }
  \label{fig:illustrate_size}
\end{figure}

We treat the colloidal dispersions as a three-component mixture of positively charged colloids (with valency $Z$ and diameter $\sigma$), monovalent counterions (with valency $-1$ and diameter $\sigma_i$) , and monovalent coions (with valency $+1$ and diameter $\sigma_i$); thus, we consider colloidal particles suspended in a 1:1 electrolyte. Throughout we set $\sigma_i=\sigma/20$ in this study. In  Fig. \ref{fig:illustrate_size}(a), these three species are illustrated. We consider additive interactions and write $\sigma_{kl}=(\sigma_k+\sigma_l)/2$ for every pair $k$ and $l$. 

The simulations are performed in a fixed  box of volume $V=L^3$ with periodic boundary conditions, where we denote the number of colloids by $N$ and the number of counter- and co-ions by $N_-$ and $N_+$, respectively. To ensure global charge neutrality, we impose $Z N + N_{+}=N_{-}$ for all configurations, both in the canonical ensemble (where the particle numbers of all species are fixed) and in the semi-grand ensemble (where only $N$ is fixed while $N_+$ and $N_-$ can fluctuate at a fixed chemical potential $\mu$ of the salt). The packing fraction of the colloids is denoted by $\eta=(\pi/6)\sigma^3 N/V$. In the salt-free limit, where $\beta\mu\rightarrow-\infty$ such that the number of co-ions vanishes ($N_+=0$), the dispersion is fully characterized by the dimensionless temperature-like variable $\sigma/\lambda_B$, the colloid valency $Z$, and the packing fraction $\eta$, where we recall that we consider monovalent ions and fix the size ratio to $\sigma_i=\sigma/20$ throughout. In the case of finite $\beta\mu$ we characterize the ionic strength by the dimensionless combination $\kappa\sigma$, where $\kappa^{-1}$ is the Debye length of the bulk electrolyte (so $\eta=0$) in osmotic equilibrium with the dispersion (at $\eta\neq0$),  as described in Section \ref{sec:electrolyte_salt} and Ref. \citenum{terRele2025}.

\subsection{Training Data Generation} 
We use the same techniques as described in Ref. \citenum{terRele2025} to  generate the training data used to construct the effective machine-learned (ML) colloidal potentials. We perform molecular dynamics (MD) simulations of the Primitive Model, from which we obtain a $3N$-dimensional vector with components ${\bf F}^{PM}_i$ that contains the ion-averaged PM force acting on colloidal particle $i = 1,\ldots,N$ in a given colloid configuration $\{{\bf R}\}$. Each training set consists of 240 configurations sampled at 120 colloid packing fractions within the range $\eta \in [0.001, 0.45]$, while keeping all other parameters fixed. 
The size of the training set was determined through  preliminary tests, which indicated that increasing the number of configurations in the training set does not significantly affect the resulting potential or the training accuracy, as quantified by the RMSE and $R^2$ values.

To initialize a training-data-generating simulation (at a given colloid valency $Z$ and temperature $\sigma/\lambda_B$), we begin by randomly placing $N$ colloids and $N_-=ZN$ charge-neutralizing counterions in a cubic box of length $L =  V^{1/3}$, ensuring that no hard-sphere overlaps occur. The box size  $L$ is chosen such that the desired packing fraction $\eta = (\pi \sigma^3/6)N/L^3$ is obtained. All primitive-model simulations used for data generation employ a fixed number of $N=32$ colloids and are performed with the LAMMPS software package.\cite{LAMMPS} Long-range Coulomb interactions are computed using the particle-particle-particle-mesh (PPPM) Ewald summation method,\cite{Hockney1988} with periodic boundary conditions. In simulations that include salt, pairs of counterions and coions are inserted and removed using a Grand Canonical Monte Carlo (GCMC) scheme, with the chemical potential $\beta \mu$ controlling the salt concentration. Further details of the GCMC scheme are provided in Appendix \ref{sec:sup:gcmc}.

The simulations are initialised by randomly placing the colloids and counterions in the simulation box, after which the system is evolved in the $NVT$ ensemble for 10000 MD time steps. For simulations with added salt, 25000 GCMC moves are performed following this initial equilibration, after which the system is further propagated for 10000 MD steps, while performing 100 GCMC moves every 200 MD steps. Afterward, both for simulations with and without added salt, the colloids are fixed in place, and the system is evolved for an additional 40000 MD steps. Finally, the system is simulated for 800000 MD steps, during which the forces on the colloids are recorded every 200 MD steps. These sampled forces are used to compute the ion-averaged Primitive Model forces  ${\bf F}^{PM}_i$  for each colloid  configuration.  A representative simulation snapshot for colloids interacting in the strongly-coupled electrostatic regime is shown in Fig. \ref{fig:illustrate_size}(b).
This procedure is repeated for $M$ distinct colloid configurations, yielding  $3N\times M$ Cartesian  components of the ion-averaged forces acting on the colloids. 
Together with the corresponding $M$ stored colloid configurations $\{\bf{ R}\}$, this data serves as input for the force-matching linear regression algorithm.

\subsection{Machine Learning Procedure}\label{sec:ML-procedure}
The procedure used to generate an ML potential is a linear-regression method, which was thoroughly described in Ref. \citenum{terRele2025} and  summarized briefly here. The potential is expressed as a weighted sum of symmetry functions,\cite{Behler2007} with the corresponding weights  chosen to minimize the Root-Mean-Squared-Error (RMSE) between the training  forces and those predicted by the ML potential. The RMSE is defined as 
\begin{equation}\label{RMSE}
    \text{RMSE} = \sqrt{\frac{\left(\mathbf{f}^{PM}- \mathbf{f}^{ML} \right)^2}{3N M}},
\end{equation}
where $\mathbf{f}^{PM}$ represents the vector of forces obtained from the primitive-model simulations  and $\mathbf{f}^{ML}$ the forces predicted by the ML potential for the same  colloid configurations. In total, the training set consists of  $M$ configurations of $N$ colloids. 

We also characterize the accuracy of the trained potentials using the coefficient of determination $R^2$, defined as
\begin{equation}
    R^2 = 1 - \frac{\left(\mathbf{f}^{ML} - \mathbf{f}^{PM}\right)^2}{\left( \bar{\mathbf{f}}^{PM} - \mathbf{f}^{PM} \right)^2},
\end{equation}
where $\bar{\mathbf{f}}^{PM}$ is a $3NM$-dimensional vector in which every entry  equals  the mean of all components of  $\mathbf{f}^{PM}$. By construction,  $\bar{\mathbf{f}}^{PM} \cdot \mathbf{1} = \mathbf{f}^{PM} \cdot \mathbf{1}$, with $\mathbf{1}$ denoting the $3NM$-dimensional vector with all entries equal to one. 
The coefficient of determination quantifies how well  the ML forces reproduce the input forces from simulations, with $R^2 = 1$ indicating perfect agreement, corresponding to an RMSE of 0.

\subsection{Coarse-Grained Simulations using ML potentials}
Once the ML potential has been generated for a given set of system parameters $\sigma_i/\sigma$, $\sigma/\lambda_B$, $Z$, and $\kappa\sigma$, it can be employed in simulations of effective colloids-only systems for all packing fractions within the training range. These simulations contain many more colloidal particles than the $N = 32$  used in the primitive-model simulations that were employed to construct  $U^{ML}(\{\bf R\})$; in practice, we typically  simulate hundreds to  thousands of colloidal particles in the colloids-only systems.   
As with the primitive  model simulations, the simulations employing ML potentials are MD simulations performed using the LAMMPS software package,\cite{LAMMPS} where we make  use of the HDNNP-package for  implementing the many-body potentials.\cite{Singraber2019}

\section{Salt-free colloidal dispersions}\label{sec:asymmetric-electrolyte}
We begin our investigation with a system composed  solely of charged colloids (macroions) and monovalent counterions,  without any added background salt. Simulations of such salt-free systems have previously revealed gas-liquid and gas-crystal phase coexistence within suitable parameter regimes.\cite{Linse1999, Linse2000, Hynninen2007} 
Throughout this section, we fix the ion-to-colloid size ratio at $\sigma_i /\sigma = 0.05$, and examine the behavior  of this system as a function  of several dimensionless parameters, in particular the charge asymmetry $Z$, the effective temperature $\sigma/\lambda_B$, and the packing fraction $\eta$. In Section \ref{sec:electrolyte_salt}, we extend the analysis to systems with added salt by considering a finite salt chemical potential $\beta\mu$.

\subsection{Charge Asymmetry $Z=100$}\label{sec:assym100}
\begin{figure}[ht]
    \includegraphics[width=0.9\linewidth]{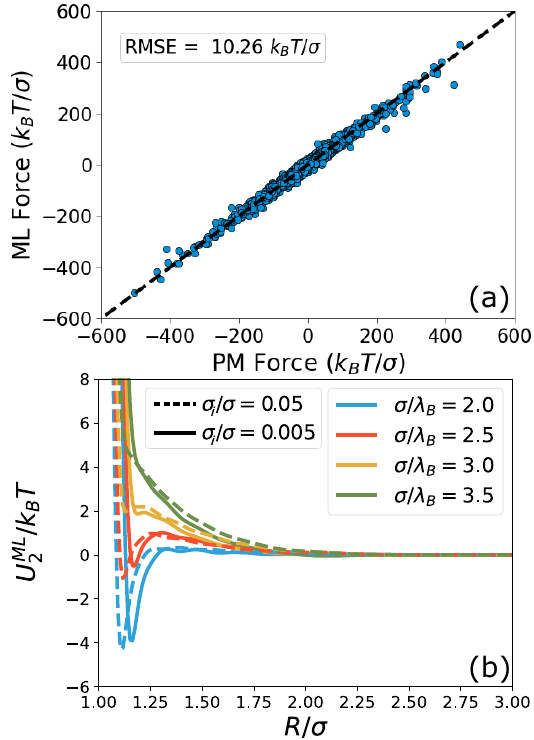}%
 \caption{(a) Parity plot comparing the Cartesian components of the effective many-body ML forces $\mathbf{F}^{ML}_{i,\alpha}$ (in units of $k_B T/\sigma$) predicted by the ML model with the corresponding PM forces $\mathbf{F}^{PM}_{i,\alpha}$ measured in primitive-model simulations for the same configurations. The ML potential describes a salt-free system of colloids with valency $Z=100$ and monovalent counterions at effective temperature $\sigma/\lambda_B = 2.5$ , with ion to colloid ratio $\sigma_i/\sigma = 0.05$  (full lines) and $\sigma_i/\sigma = 0.005$ (dashed lines). The ML model was trained on 240 configurations of 32 colloids at packing fractions in the range $\eta \in \lbrack0.001, 0.45\rbrack$.
 (b) The effective two-body ML potential $U_2^{ML}(R)$ for effective temperature $\sigma/\lambda_B \in \lbrack2.0, 3.5 \rbrack$, trained using the same procedure as the potential in plot (a).}
  \label{fig:hyn_force_compare}
\end{figure}

We first consider a system consisting of colloids with a charge number $Z = 100$ and an effective temperature in the range $\sigma/\lambda_B \in [2.0, 3.5]$, which corresponds to $Z \lambda_B /\sigma \in [16.7,50.0]$. We perform Primitive Model (PM) simulations on $N=32$ colloids and their counterions at 120 different packing fractions in the range  $\eta \in [0.001, 0.45]$. For each packing fraction $\eta$, we collect two  distinct colloid configurations $\{ \mathbf{R}\}$, and measure the ion-averaged PM forces acting on each colloid. We train a ML model using the ion-averaged PM forces obtained from $M=240$ colloid configurations, where we use a cut-off radius of $R_c = 2.5 \sigma$. The resulting ML potential for $\sigma/\lambda_B = 2.5$, constructed using $D=20$ symmetry functions, achieved a root-mean-square error $\text{RMSE} = 10.26 k_B T/\sigma$ and a coefficient of determination of $R^2 = 0.979$. In Fig. \ref{fig:hyn_force_compare}(a) we show a parity plot comparing the Cartesian components of the ML-predicted effective many-body  forces $\mathbf{F}^{ML}_{i,\alpha}$ with the corresponding PM forces $\mathbf{F}^{PM}_{i,\alpha}$ measured in the test configurations obtained from the primitive-model simulations. We obtain comparable accuracy for other values of $\sigma/\lambda_B$.

Next, we evaluate the effective ML pair potential $U^{ML}_2(R)$, predicted by our ML model  for a system containing  only two colloids at center-to-center distance $R$. In Fig. \ref{fig:hyn_force_compare}(b), we present $U^{ML}_2(R)$, which is obtained from training on Primitive Model simulation data. Interestingly, we observe that the effective ML potentials exhibit characteristic features of electrostatic interactions in the strong-coupling regime. Specifically, at the highest effective temperature, $\sigma/\lambda_B = 3.5$, we observe that the effective two-body ML potential remains purely repulsive, consistent with the predictions from Poisson-Boltzmann (PB) theory. However, as the effective temperature decreases, an attractive potential well forms at small colloid distances $R$. Initially, this well appears as a shallow dip within an otherwise repulsive potential. Upon decreasing the temperature further, the attraction  becomes more pronounced and  eventually dominates, suppressing  all repulsive features for distances $R \geq 1.2\sigma$ at the lowest effective temperature $\sigma/\lambda_B=2.0$. These observations align with the attractive well previously reported in both simulations\cite{Hynninen2007, Lin2021} and experiments.\cite{Stelmakh2019}  Additionally, we find that varying the ion-to-colloid size ratio has only a limited effect on the resulting ML potentials. In Fig. 2, which shows ML potentials at four different temperatures $\sigma/\lambda_B$, we observe that for the much smaller ion-to-colloid size ratio $\sigma_i/\sigma = 0.005$, the repulsive bare colloid charge interactions set in  at relatively smaller separations compared to the potential for $\sigma_i/\sigma = 0.05$. However, the overall character of the potential is fully preserved when reducing the ion size from small to even smaller. In particular, the depth and range of the attractive well at short separations, as well as the strength and range of the repulsion at the highest temperature, remain essentially  unchanged. This comparison supports and justifies our choice of  $\sigma_i/\sigma = 0.05$ throughout the present work.

\begin{figure}[ht]
    \includegraphics[width=0.9\linewidth]{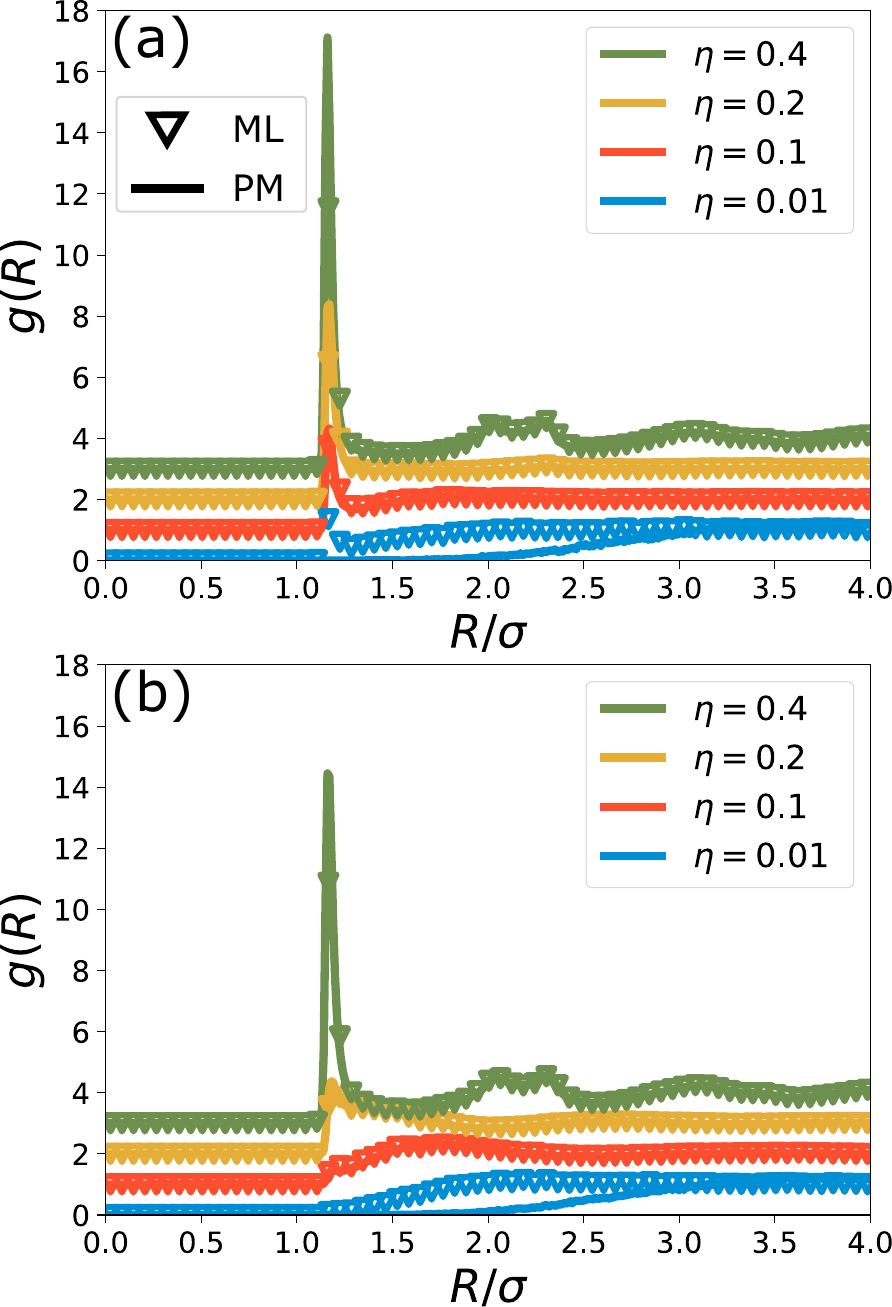}%
  \caption{
  Radial distribution function $g(R)$ as obtained from primitive-model simulations (lines) and from coarse-grained simulations using ML potentials (triangles) for a salt-free colloidal dispersion with $Z=100$ and ion-to-colloid size ratio $\sigma_i/\sigma=0.05$ as described in Section \ref{sec:assym100}
  at several  packing fractions $\eta$ (see labels), for effective temperatures (a) $\sigma/\lambda_B = 2.5$ and (b) $\sigma/\lambda_B = 3.0$. The ML model is trained on configurations with packing fractions in the range of $\eta \in [0.001, 0.45]$. } 
  \label{fig:hyn_RDf_compare}
\end{figure}

We also evaluate the performance of the effective ML potentials by comparing the colloid-colloid radial distribution functions $g(R)$, obtained from primitive-model simulations and from molecular dynamics simulations using the effective ML potential, at various packing fractions within the range  $\eta \in [0.01, 0.4]$.  In Fig. \ref{fig:hyn_RDf_compare}, we plot the $g(R)$  for effective temperatures  equal to (a) $\sigma/\lambda_B =2.5$ and (b) 3.0. From Fig. \ref{fig:hyn_RDf_compare} we clearly observe that simulations using the effective ML potential accurately reproduce the radial distribution functions of the primitive-model simulations at both temperatures, especially for $\eta\geq 0.1$, where even the significant first peak is captured quantitatively. 
However, for both temperatures, the agreement is significantly worse at  $\eta=0.01$, where the range of the ML-based $g(R)$ is too short, and it reaches its first peak at a much smaller distance than that of the PM-based $g(R)$.
We expect that retraining at lower packing fractions could remedy this shortcoming; however, we did not pursue this further, as   
the $g(R)$'s  from simulations using the ML potential and the Primitive Model show overall satisfactory agreement.

\begin{figure*}[ht]
    \includegraphics[width=0.9\linewidth]{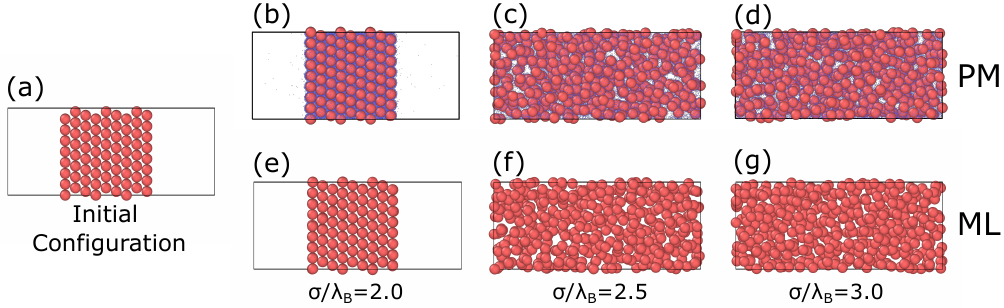}%
  \caption{Typical configurations from primitive-model simulations (top row) and coarse-grained simulations using  ML potentials (bottom row) for a salt-free colloidal suspension  ($\exp{[\beta\mu/2]} \rightarrow0$) with $Z = 100$ and ion-to-colloid size ratio $\sigma_i/\sigma=0.05$. (a) Initial configuration in all simulations is a face-centered cubic crystal phase adjacent to a vacuum. (b), (c), and (d) Final configurations after equilibration from primitive-model simulations at effective temperatures $\sigma/\lambda_B = 2.0, 2.5,$ and $ 3.0$, respectively. (e), (f), and (g) Final configurations after equilibration from simulations using the ML potentials at the same temperatures. 
  }
  \label{fig:FCC_coex_overview}
\end{figure*}

In addition to the fluid structure, we also compare aspects of the phase behavior, specifically regarding the presence of phase separation. 
For several effective temperatures $\sigma/\lambda_B$, we perform direct-coexistence simulations using both the Primitive Model and the ML potential. In each case, the system is initialized as a colloidal face-centered cubic crystal phase positioned next to a vacuum. After equilibration, we assess the equilibrium state of the system. The crystal packing fraction, $\eta = 0.45$, is carefully chosen such that neighboring particles initially reside at the minimum of their mutual two-body potential. In Fig. \ref{fig:FCC_coex_overview}, we show snapshots of the time evolution of the simulations with the ML potential and of the primitive-model simulations for three effective temperatures, revealing a clear similarity between configurations from the primitive-model simulations (upper row) and the corresponding simulations with the ML potential (lower row). For $\sigma/\lambda_B = 2.0$, the initial crystal remains intact, indicating mechanical stability of the crystal and the possibility of a phase coexistence of a colloidal crystal and a very dilute colloidal gas. By contrast, for $\sigma/\lambda_B = 3.0$ the crystal quickly melts into a homogeneous fluid, while for the intermediate effective temperature $\sigma/\lambda_B = 2.5$, we find for both the simulations with the Primitive Model and the ML potential that the crystal melts slowly and eventually becomes a homogeneous fluid. Additionally, both the ML and PM simulations showed spontaneous phase separation at $\sigma/\lambda_B = 2.0$, lending strong support for the presence of a spinodal instability at sufficiently low temperatures. 

Altogether, the ML potential is accurate in reproducing results from primitive-model simulations for a salt-free colloidal dispersion with a charge asymmetry $Z = 100$. Already for colloid charge number $Z = 100$, the simulations of Fig.~\ref{fig:FCC_coex_overview} with the the ML potential are more than 10 times faster than the primitive-model simulations. In the following section, we use the same method to train ML potentials at higher charge numbers in order to predict the phase behavior of salt-free colloidal systems with colloid valencies $200 \leq Z \leq 1000$.

\subsection{Higher Colloid Valency }\label{sec:assymhigh}
Here we extend the investigation of the salt-free suspension with colloid valency $Z=100$ of Section \ref{sec:assym100} to higher colloid valencies in the range  $Z\in[200,1000]$.  Our focus is on identifying the effective temperature $\sigma/\lambda_B$ below which stable crystals emerge, indicating phase coexistence with a dilute colloidal gas phase whose packing fraction  is several orders of magnitude lower than that of the crystal phase. 

Since the work of Van der Waals, it has been known that the coexistence of a dilute and a much denser phase requires cohesive energy that  stabilizes the dense (liquid or crystalline) phase, while the dilute gas phase is stabilized by its high entropy per particle. In the case of highly charged colloids at low (or even zero) salinity this is a contentious issue as the required cohesive energy suggests ``like-charge attraction'', a phenomenon that is at odds with the pairwise repulsive screened-Coulomb interaction predicted by DLVO theory (at least for index-matched suspensions in which attractive dispersion forces are negligible). As mentioned already in the introduction, experimental observations of (alleged) like-charge attractions during the past decades\cite{Tata1992, Ito1994, Tata1997, Larsen1997} have triggered several theoretical explanations, however, so far without a comparative test by direct microscopic simulations of an explicit many-colloid system, since the required computer power is insurmountable. Even in the case of the Primitive Model of interest here (so without explicit water molecules), the high colloid valency $Z\simeq10^3-10^4$ of the experiments has so far not been addressed by direct simulations of many-colloid systems, especially not at finite background salt concentrations. 
We again restrict ourselves to generating potentials for monovalent ions, since the dominant counterion species in the experimentally studied colloidal suspensions were monovalent as well (either $K^{+}$\cite{Monovoukas1989} or $H^+$ \cite{Tata1997, Larsen1997, Gomez2009}).

With our machine learning approach, we can extend the computational study toward  experimentally relevant  charge valencies and investigate the mechanisms behind  like-charge attraction. Once we have trained an ML potential for a given parameter set with a high charge valency, we can perform large-scale simulations, enabling the investigation of phase coexistence at charge numbers that were previously computationally inaccessible. 
Here we compare two candidate explanations for the observed phase coexistence: one involving non-pairwise many-body interactions as encoded by ``volume terms'' and the other attributing the effect to  correlated fluctuations in the low-temperature, strong-coupling regime.

\begin{figure*}[!ht]
    \includegraphics[width=0.85\linewidth]{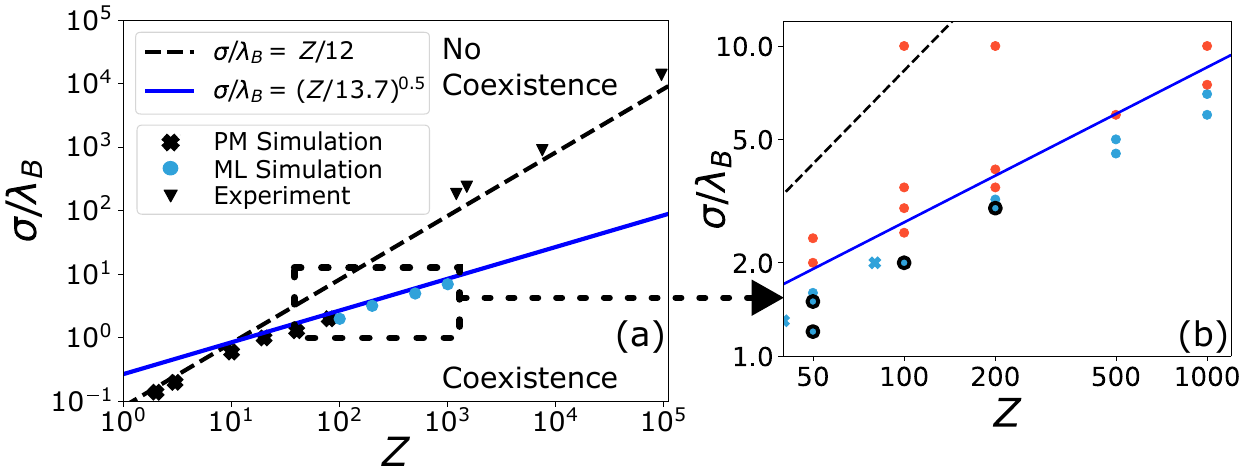}%
 \caption{Double logarithmic representation of the colloid valency $Z$ and effective temperature $\sigma/\lambda_B$ plane for a binary (salt-free) colloid-counterion mixture at ion-to-colloid size ratio $\sigma_i/\sigma=0.05$. The full plane in (a) shows the highest temperatures at which our ML simulations exhibit gas-crystal coexistence (blue dots) and includes literature results (black symbols, lines), and the zoom for  $Z\in[50,1000]$ in (b) shows the presence (blue dots), with black rings indicating spontaneous phase separation, and absence (red dots) of gas-crystal phase coexistence as obtained from simulations using the ML potential. The crosses in (a) correspond to  critical temperatures found from simulations of asymmetric electrolytes for $Z = 2,3$ \cite{Panagiotopoulos2002}  and for $Z = 10, 20, 40, 80$,\cite{Linse2000} while the  triangles indicate experimentally observed coexistence or like-charge attraction in colloidal suspensions at low salinity with increasing $Z$ (Refs.  \citenum{Monovoukas1989}, \citenum{Tata1997}, \citenum{Larsen1997}, \citenum{Gomez2009}). 
   Note that the charge valencies reported in the experiments inherently account for charge regulation, as they were determined from conductivity measurements.}
 The two lines are literature estimates of the ``critical'' effective temperature separating  regimes with and without gas-crystal coexistence, the strong-coupling result (blue) given by $\sigma/\lambda_B=(Z/13.7)^{0.5}$ from MC simulations,\cite{Hynninen2007} and the weak-coupling Poisson-Boltzmann prediction $\sigma/\lambda_B=Z/12$ (black dashed).\cite{Zoetekouw2006prl}
  \label{fig:coex_overview_tot}
\end{figure*}

We generate effective ML potentials for colloid charges in the range $Z \in [50, 1000]$ for various effective temperatures $\sigma/\lambda_B$, using the same protocols as for the $Z=100$ case described in  Section \ref{sec:assym100}. All simulations are performed at a fixed ion-to-colloid size ratio $\sigma_i/\sigma = 0.05$ and in the salt-free limit,  $\exp{[\beta\mu/2]} \rightarrow 0$. We then employ these ML potentials in simulations where a  face-centered cubic colloidal crystal is initially placed adjacent to a vacuum,  and we examine whether the crystal melts upon equilibration, following the same procedure as in the simulations with the ML potentials shown in Fig. \ref{fig:FCC_coex_overview}. 
The effective ML potential for charge valency $Z = 50$ was already presented in Ref. \citenum{terRele2025}, although without a discussion on the phase separation that occurs. 
The packing fraction $\eta$ of this crystal is again chosen such that neighbouring particles reside at the minimum of their mutual two-body ML potential $U_2^{ML}$. 
Furthermore, we employ the ML potentials in simulations of homogeneous colloidal fluids, to gauge if these ML potentials result in spontaneous clustering.  
In the $Z -\sigma/\lambda_B$ plane shown on a double logarithmic scale in Fig. \ref{fig:coex_overview_tot}, the blue dots obtained from simulations with the ML potentials for $Z\in[100,1000]$ in panel (a) indicate the highest effective temperatures $\sigma/\lambda_B$ at which gas-crystal coexistence was observed. In the zoomed-in view of panel (b), blue symbols mark state points where gas-crystal coexistence occurs, whereas red symbols indicate state points where the crystal melts entirely, and no phase coexistence persists. In addition, the black ringed blue dots indicate where phase separation occurs spontaneously in ML potential simulations initialized as a homogeneous colloidal fluid. 

Interestingly, our ML prediction for the crossover between  coexistence and melting in the range  $Z\in[50,1000]$ aligns well with the blue line representing $\sigma/\lambda_B = (Z/13.7)^{0.5}$, which was proposed by Hynninen \textit{et al.} based on simulations of critical points for asymmetric electrolytes in the lower-charge regime $10<Z<80$.\cite{Hynninen2007} The first appearance of spontaneous phase separations follows a trend similar to the Hynninen prediction, although the temperature at which this occurs is slightly lower than predicted by Hynninen and the crystal-melting temperature. Although we observe mechanical stability of the crystal phase for $Z=500$ and $Z=1000$, we do not observe spontaneous phase separation for any ML potential trained on colloids with the same colloid valencies. 
At these high charge valencies, the two-body potential contains a large repulsive barrier, hindering spontaneous crystallization. Additionally, the crystal stability depends strongly on the attractive three-body interactions.
The black crosses obtained from primitive-model simulations\cite{Linse2000, Panagiotopoulos2002} in (a) align with these simulated points, including those for small valencies $Z=2,3$, which lie much closer to the black dashed line $\sigma/\lambda_B=Z/12$---as predicted by the Poisson-Boltzmann-based volume-term theory of Ref. \citenum{Zoetekouw2006prl}. This theory was designed to explain experimentally observed gas-crystal coexistence in colloidal systems, shown in (a) by black triangles for colloid valencies in the range $Z\in[10^3,10^5]$. 

Although the volume-term theory results appears to provide a reasonable interpolation between the low-$Z$ coexistence regime of electrolytes and the high-$Z$ coexistence regime of colloidal dispersions,\cite{RvR2000} our ML results indicate that, at least within salt-free primitive-model  descriptions without explicit water, phase coexistence at high-$Z$  requires a lower effective temperature, given by $\sigma/\lambda_B\leq(Z/13.7)^{0.5}$. Note that this low-temperature regime corresponds to a high Coulomb coupling parameter $\Xi \equiv 2(\lambda_B/\sigma)^2 Z \geq 13.7$, which therefore corresponds to the strong-coupling regime.\cite{Punkkinen2008} 

Thus, the attractive interactions characteristic of the electrostatic strong-coupling regime appear to provide the cohesive energy that drives and stabilizes the broad phase coexistence observed in the Primitive Model across a wide range of colloid charges. Moreover, our simulations using the ML potentials do not support the earlier suggestion by Zoete\-kouw \textit{et al.} that the phase separation observed at low charges $Z$ and low temperatures $\sigma/\lambda_B$ in primitive-model simulations of asymmetric electrolytes can be directly extrapolated to the high-charge,  high-temperature colloidal regime. The experimental observation of like-charge attraction in colloidal systems at relatively high effective temperatures $\sigma/\lambda_B\simeq Z/12$ for $Z\in[10^3,10^4]$, therefore requires an alternative explanation. 
We are currently constrained by the computational cost of generating training data for charge valencies $Z>10^3$ using PM simulations. To further extend our conclusions, one  would need to move beyond the PM framework and consider, for example,  corrected Poisson-Boltzmann theory approaches that  incorporate ion-correlations.\cite{Guan2016, Lesniewska2025} However, such approaches would still be unable to capture  strong-coupling  attractions between like-charged particles, which  require an explicit treatment of individual ions. 
Alternatively, the ML-predictions might be extended to higher charge valencies by employing  lattice-based PM simulations,\cite{Diehl2005, Hynninen2007} which could offer a computationally more efficient route while retaining an explicit description of ionic degrees of freedom.

\section{The Effect of Salt}\label{sec:electrolyte_salt}
In Sections~\ref{sec:assym100} and \ref{sec:assymhigh}, all simulations were performed for salt-free systems, i.e. without coions. In this section, we investigate the effect of added salt on the effective ML potential $U_2^{ML}$ between colloids, on the resulting radial distribution function, and on the emergence of gas-crystal phase coexistence. To generate training data for these potentials, salt is added to and removed from the simulation box using the  Grand Canonical Monte Carlo protocol described in Appendix \ref{sec:sup:gcmc}, where the (average) salt concentration is controlled by a finite chemical potential $\mu$ of salt pairs. Rather than characterizing the ionic strength by the ion activity $\exp(\beta\mu/2)$, we use the combination $\kappa\sigma$ where $\kappa^{-1}=(4\pi\lambda_Bc_s)^{-1/2}$ represents the Debye screening length of a 1:1 salt reservoir (without colloids) at Bjerrum length $\lambda_B$ and at the $\beta\mu$-dependent ion concentration $c_s$; this implies that the reservoir is in osmotic equilibrium with the suspension.\cite{terRele2025} 

We begin by investigating systems with colloid valency $Z=100$ and ion-to-colloid size ratio $\sigma_i/\sigma=0.05$, at the two effective temperatures $\sigma/\lambda_B = 2.5$ and  $\sigma/\lambda_B = 3.0$. 

\begin{figure}[ht]
\hspace*{-0.5cm}
\includegraphics[width=0.9\linewidth]{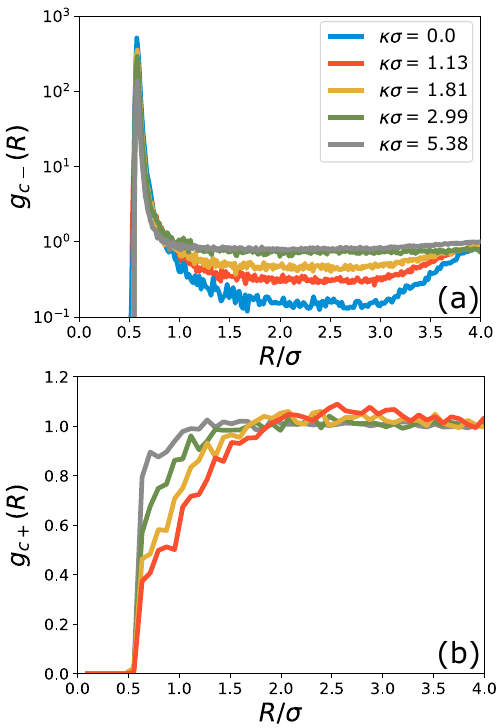}
    \caption{Colloid-counterion and colloid-coion radial distribution functions $g_{c-}(R)$ (a) and $g_{c+}(R)$ (b), respectively,  obtained from primitive-model simulations for several screening parameters $\kappa\sigma$ (colors),  for effective temperature $\sigma/\lambda_B=2.5$, with colloid valency $Z = 100$ and ion-to-colloid size ratio $\sigma_i/\sigma = 0.05$ in all cases. }
  \label{fig:ionrdf_Ts2.5}
\end{figure}

In Fig. \ref{fig:ionrdf_Ts2.5}(a) we show the colloid-counterion radial distribution functions $g_{c-}(R)$ and $g_{c+}(R)$, respectively,  for several values of $\kappa\sigma\in[0,6.0]$ (colors) at an effective temperature  $\sigma/\lambda_B = 2.5$. The prominent peak of $g_{c-}(R)$ at quasi-contact distance $R\simeq 0.55 \sigma$ indicates that counterions are strongly bound to the oppositely charged colloid surface, characteristic of the strong-coupling regime. As $\kappa \sigma$ increases, this peak becomes less pronounced because a larger fraction of  counterions is no longer  attached to a colloid and can instead  move rather freely in solution.  The colloid-coion pair distribution $g_{c+}(R)$ in (b) exhibits pronounced depletion near contact, with the depletion length decreasing as $\kappa\sigma$ increases---qualitatively in line with conventional Poisson-Boltzmann theory.    
We find very comparable counterion- and coion-distributions around a colloid for effective temperature $\sigma/\lambda_B = 3.0$ (not shown).

\begin{figure*}[ht]
    \includegraphics[width=0.72\linewidth]{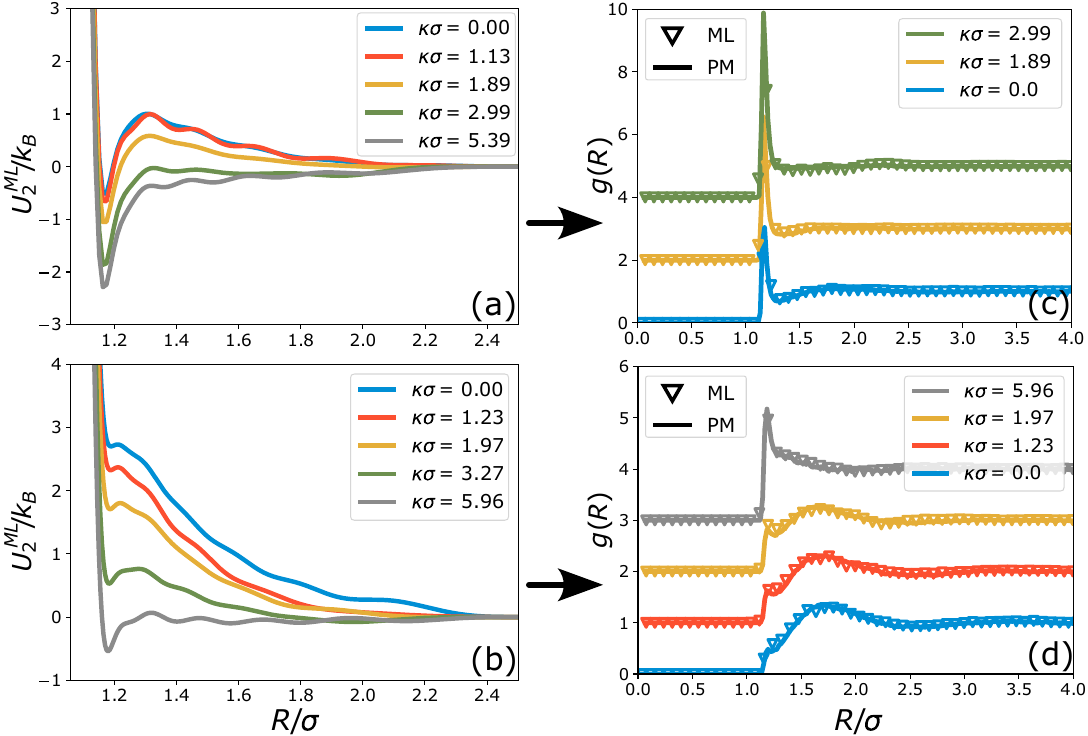}%
  \caption{ Two-body ML potentials $U_{2}^{ML}(R)$ for charged colloids with  charge valency $Z = 100$ and ion-to-colloid size ratio $\sigma_i/\sigma = 0.05$  at effective temperatures (a) $\sigma/\lambda_B = 2.5$ and (b) $\sigma/\lambda_B = 3.0$,  for varying screening parameters $\kappa \sigma$ (as labeled).
  The corresponding (shifted) colloid-colloid radial distribution function $g(R)$ at packing fraction $\eta=0.1$,   as obtained from PM simulations (full lines) and ML potential simulations (triangles), are shown for (c) $\sigma/\lambda_B = 2.5$ and (d) $\sigma/\lambda_B = 3.0$.}
  \label{fig:hyn_T3_saltpot}
\end{figure*}

In Fig.~\ref{fig:hyn_T3_saltpot}(a) and (b) we present the resulting ML pair potential $U_2^{ML}(R)$ at these two temperatures, respectively, for a range of screening parameters $\kappa\sigma\in[0,6.0]$ (colors), including the potentials in the salt-free limit $\kappa\sigma=0$ that were already presented in Fig.~\ref{fig:hyn_force_compare}(b). The effect of adding salt is profound as it enhances the attractions in both cases: at the lower temperature in (a) the potential barrier at $R=1.3\sigma$ disappears completely as $\kappa\sigma$ increases while the attractive well at $R=1.2\sigma$ deepens to form a fully attractive potential. At the higher temperature in (b) the long-ranged, purely repulsive potential at $\kappa\sigma=0$ disappears and develops a weakly attractive well at quasi-contact for $\kappa\sigma=5.96$. 

The increased attractions in $U_2^{ML}(R)$ for increasing $\kappa\sigma$ has also repercussions for the resulting colloid-colloid radial distribution function $g(R)$.  
In Fig. \ref{fig:hyn_T3_saltpot}(c), we plot the (shifted) $g(R)$ for $\sigma/\lambda_B = 2.5$ at packing fraction $\eta = 0.1$  for various values of $\kappa\sigma$, both from colloids-only simulations with the many-body ML potential $U^{ML}(\{\bf R\})$ (solid lines) and from direct PM simulations (symbols), where the latter are computationally feasible due to the relatively low valency  $Z=100$. We observe good agreement between the ML- and PM-based radial  distribution functions, with $g(R)$ dominated by a pronounced peak at the colloid-colloid quasi-contact distance $R = 1.2 \sigma$ for all screening parameters $\kappa \sigma$ considered, while the other structural features in (c) are insignificant. Apart from the higher quasi-contact peak at higher $\kappa\sigma$, there appears to be no qualitative  change in the structure.  
By contrast, the structure at the higher effective temperature $\sigma/\lambda_B = 3.0$ does change significantly with increasing $\kappa \sigma$. The (shifted) radial distribution functions $g(R)$ for $\sigma/\lambda_B =  3.0$ are plotted in Fig.~\ref{fig:hyn_T3_saltpot}(d) for various values of $\kappa\sigma$ at a colloid packing fraction of $\eta = 0.1$, again showing excellent agreement of ML- and PM-based simulations.  The change of $g(R)$ with increasing $\kappa\sigma$ is well-captured, in particular the shift from  a small primary peak at  center-to-center distance $R = 1.6 \sigma$ for $0\leq\kappa \sigma \leq  1.97$ to a pronounced primary peak at $R=1.2\sigma$ for $\kappa\sigma=5.96$.

\begin{figure}[ht!]
    \includegraphics[width=0.95\linewidth]{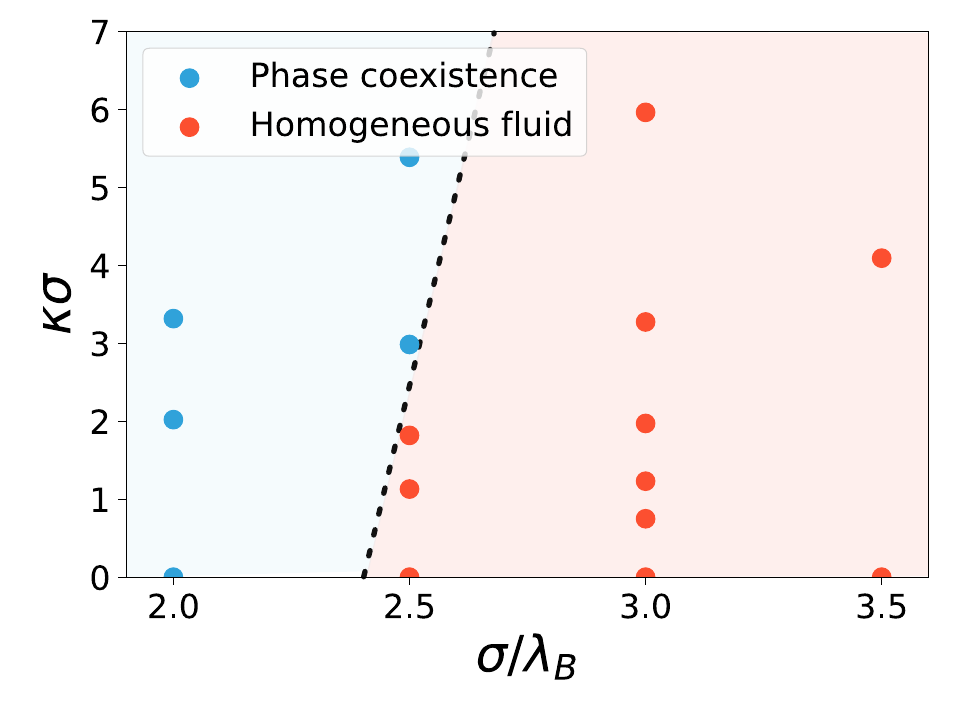}%
  \caption{State diagram in the temperature $\sigma/\lambda_B$ - screening parameter $\kappa\sigma$ representation, showing the  regimes in which gas-liquid or gas-solid phase coexistence is observed (blue dots) and  where homogeneous fluid states occur (orange dots). Results are based on   simulations using ML potentials generated for colloids with valency $Z = 100$ and  ion-to-colloid size ratio $\sigma_i / \sigma = 0.05$.  Each dot represents a simulation performed with a separately trained ML potential. The dashed line is a guide to the eye and should not be taken as the precise boundary between the two phases.}
  \label{fig:salt_coex_over_Z100}
\end{figure}

As before we investigate the (mechanical) stability of initially prepared fcc crystals adjacent to a vacuum for $Z=100$ at several temperatures $\sigma/\lambda_B\geq 2$. However, we now also consider non-zero salt concentrations and  use only the ML many-body potential $U^{ML}(\{\bf R\})$, thus without performing additional (and computationally expensive) direct PM simulations for comparison. In the plane spanned by temperature and salt concentration, represented in  Fig.~\ref{fig:salt_coex_over_Z100} by $\sigma/\lambda_B\in[2,3.5]$ and screening parameter $\kappa\sigma\in[0,7.0]$, we distinguish state points with a (mechanically) stable crystal-gas coexistence (blue dots) at low temperatures from those corresponding to a homogeneous fluid phase (orange dots) at higher temperatures. However, perhaps surprisingly, we also see that a  finite salt concentration promotes the stability of the initial crystals, since the approximate melting line (dashed) shifts to higher temperatures for increasing $\kappa\sigma$. For $\lambda_B/\sigma=2.5$, for instance, Fig.~\ref{fig:salt_coex_over_Z100} shows that the crystals melt for $\kappa\sigma\lesssim2.0$ (in agreement with their melting under salt-free conditions shown in Fig.~\ref{fig:coex_overview_tot}), whereas for $\kappa\sigma\gtrsim 3.0$ the crystals remain stable at this temperature. For the higher temperature $\sigma/\lambda_B=3.0$ we find unstable (melting) crystals for all $\kappa\sigma$ we investigated in Fig.~\ref{fig:salt_coex_over_Z100}, despite the substantial effect of $\kappa\sigma$ on $g(R)$ as shown in Fig.~\ref{fig:hyn_T3_saltpot}(d). In the Appendix D, we present this state diagram over an extended temperature range $\sigma/\lambda_B$ for $Z=100$ and $Z=200$. In this strong-coupling regime, the addition of salt promotes like-charged colloid attractions, which in turn mechanically stabilizes crystals of these like-charged colloids. However, this effect only results in a change in phase behavior for a relatively small range of temperatures.

\section{Conclusion}
In summary, we have applied a machine-learning (ML) framework to construct effective interaction  potentials for charged spherical colloids of fixed valency $Z$ suspended in a 1:1 electrolyte with Bjerrum length $\lambda_B$. The training data used to obtain these effective ML potentials stem from simulations of globally charge-neutral two- and three-component primitive-model systems consisting  of colloidal particles, counterions, and (in the case of added salt) co-ions. Our focus has not only been on the dependence of the ML many-body interactions on the temperature-like variable $\sigma/\lambda_B$ and the reservoir screening parameter $\kappa\sigma$ at a fixed ion-colloid diameter ratio $\sigma_i/\sigma=0.05$, but also on the resulting radial distribution functions at colloid packing fraction $\eta$ and on aspects of gas-crystal coexistence.

The initial focus was on salt-free dispersions ($\kappa\sigma=0$) with colloid valency $Z=100$, for which we generated ML potentials applicable over a range of $\eta$ values and several choices of $\sigma/\lambda_B$. Upon lowering $\sigma/\lambda_B$ from 3.5 to 2, the ML pair potential in the dilute limit was found to develop (like-charge) attractions, which we identified to stem from correlated fluctuations rather than from ``volume-terms''. Applying the ML many-body potential to a judiciously prepared face-centered cubic crystal in contact with a supernatant void, we observed melting at high-temperature ($\sigma/\lambda_B\geq 2.5$)  but mechanical stability of the crystal at low-temperature ($\sigma/\lambda_B=2.0$). Additionally, we observed spontaneous gas-crystal phase separation  when we initialized the system in a disordered fluid phase in both PM simulations and ML potential simulations. In all cases, we found excellent agreement between ML and PM results, which encouraged us to extend this study to larger valencies, up to $Z=1000$, where direct PM simulations become prohibitively time consuming. Based on ML simulations for system sizes unattainable for PM simulations, we  again found high-temperature melting and low-temperature mechanical stability of carefully prepared initial configurations that represent gas-crystal coexistence. For $Z\in[50,200]$, the ML simulations also showed spontaneous gas-crystal phase separation, providing further support of a spinodal instability at sufficiently low temperatures $\sigma/\lambda_B$. Interestingly,  the resulting melting line for $Z\in[100,1000]$ closely follows  $\sigma/\lambda_B=(Z/13.7)^{1/2}$, consistent with earlier  PM simulations of $Z:1$ electrolytes for $Z\in[10,100]$. This salt-free melting line lies, however, at much lower temperatures (i.e. at much higher Coulomb coupling) than earlier experimental observations of colloidal gas-crystal coexistence at low (although non-zero) salinity, see Fig.~\ref{fig:coex_overview_tot}. In addition, this figure shows that a linear extrapolation of the Debye-H\"{u}ckel-type transition line $\sigma/\lambda_B=Z/12$ (dashed black line), which is appropriate for simple electrolytes (where $Z$ is of order unity), all the way up to the colloidal regime (where $Z\gtrsim 10^4$) is unreliable. For $Z \gtrsim 10$, the actual melting line in Fig.~\ref{fig:coex_overview_tot} ``bends'' downward to lower temperatures, closely following $\sigma/\lambda_B=(Z/13.7)^{1/2}$ (blue solid line). 

Next we studied the effect of added salt by constructing ML potentials for non-zero screening $0<\kappa\sigma\lesssim 8$, focusing on $Z=100$ in the temperature regime $\sigma/\lambda_B \in [2,3.5]$ where the salt-free dispersion was found to exhibit a crossover from stable to melting crystals. The emerging general picture is that the addition of salt, at least in this parameter regime, promotes (pairwise) attractions that become stronger at lower temperatures for fixed $\kappa\sigma$. Since our model does not include any dispersion forces,  the origin of the attractions in this strong-coupling regime stems from correlated ion-ion fluctuations in the vicinity of the colloid surfaces from which the ions can escape somewhat upon the addition of salt.  The addition of salt also increases the stability of carefully prepared colloidal crystals that are in contact with a colloid-poor (gas) phase, in the sense that the melting temperature of the crystal increases with increasing $\kappa\sigma$ as shown in Fig.~\ref{fig:salt_coex_over_Z100}. However, the coexistence parameter region presented in Fig.~\ref{fig:coex_overview_tot} is only marginally extended by adding salt.

The Primitive Model employed in this work captures many of the key features governing the interactions between charged suspended particles. However, it does not incorporate all effects that may influence like-charged colloidal  interactions, such as solvent polarization near the colloid surface. For instance, to account for  polarization-induced electrosolvation forces, as discussed in Refs.\citenum{Kubincov2020, Wang2024,Wang2025}, one would require  models that simultaneously treat colloidal charge regulation, ion screening, and solvent polarisation. 
However, such a  set-up is computationally too intensive to fall within the scope of the present work, and will instead be the subject of  future research.

The literature predictions of room-temperature phase separation in aqueous dispersions at low ionic strength, based on volume-term theories, are  not supported by our ML-based simulation results. On the premise of the reliability of our well-tested ML potentials, we are led to the conclusion that the Primitive Model in the weak-coupling regime does not exhibit gas-crystal coexistence. It appears that the linearised Poisson-Boltzmann theory that underlies the volume terms, which do predict gas-crystal coexistence in this parameter regime, overestimates the cohesive energy that drives demixing and does not properly capture the subtle free-energy balance that governs phase separation. This finding would also imply that the experimental observations of like-charge attraction and phase separation in room-temperature colloidal  dispersions with 1:1 electrolytes require physical ingredients beyond those included in the Primitive Model. 
Possible factors beyond the Primitive Model  include an explicit and polarisable solvent,\cite{Kubincov2020} solvation and hydration effects of dissolved ions,\cite{Bonthuis2012} and charge regulation. The standard PB framework has  recently been extended to include the effects of ion size, dielectric decrement, and ion correlations on colloidal interactions.\cite{Lesniewska2025} While finite ion size and ion correlations are already captured by the Primitive Model, and can thus be compared directly to PM results, dielectric decrement,  arising from  changes in ion concentration, is not included in the PM, although its effect on colloid interactions is expected to be  minimal. The salt concentrations studied in this work are low ($c_s\sim 5\mu \text{M}$), and assuming a typical decrement value of $\gamma \approx 10 \text{M}^{-1}$,\cite{BenYaakov2011}  the dielectric constant $\epsilon$ changes negligibly.  
The renormalized Jellium model provides an alternative description for the interactions between colloids, by treating most colloidal charges as only contributing to a uniform background charge. This model could provide further insights into  colloidal interactions at low salt concentrations.\cite{Trizac2004, dosSantos2026}

In conclusion, on the basis of Primitive-Model simulations of highly charged colloids with valencies $Z\in[100,1000]$ and monovalent ions, we have machine-learned many-body potentials for the corresponding effective colloids-only system over a substantial range of system parameters, including added salt concentration and size and charge of the colloidal particles. Using these ML potentials in simulations of ${\cal O}(10^2-10^3)$ colloidal particles, we were able to explore the structural and thermodynamic properties in a hitherto largely inaccessible parameter regime. Our ML results agree with direct PM simulations in regimes where the comparison can be made. Nevertheless, this study still covers only a relatively small part of the huge parameter space of the PM, in particular $Z\in[100,1000]$, $\sigma/\lambda_B\in[2,4]$, $\kappa\sigma \in [0,6]$, and  $\eta\in [0.001,0.4]$. Throughout, the ion-to-colloid radius was fixed at $\sigma_i/\sigma=0.05$ (such that $\sigma_i/\lambda_B\in[0.1,0.2]$) and all ions were taken to be monovalent. We hope that our study can serve  as a stepping stone toward, for instance, modeling actual aqueous colloidal dispersions of micron-sized particles at low salinity, where $\lambda_B\simeq1\text{nm}$, $\sigma\simeq1\,\mu\text{m}$, $Z\simeq10^4-10^5$, and $\kappa\sigma\simeq 1-10$. This regime implies much higher valencies, a temperature as high as $\sigma/\lambda_B\simeq10^3$, and a size ratio as small as $\sigma_i/\sigma\simeq 10^{-3}$. The required $\kappa\sigma$-regime is quite similar to that in the present case, however reaching this regime in PM simulations to machine-learn the potential requires a huge number of salt ions, even at $\kappa\sigma=0$. Extrapolations from the presently used parameter set therefore seems to be a more viable approach. Other challenges involve extensions to, for instance, charge-regulating colloids at a fixed zeta-potential or with a given titration charge and pK to account for protonation-deprotonation equilibria. For all these cases, the ML approach we explored here seems to be applicable as long as the microscopic simulations can be performed.

\begin{acknowledgments}
The authors thank Tim Veenstra and Gerardo Campos-Villalobos for many useful discussions. T.t.R. and M.D. acknowledge funding from the European Research Council (ERC) under the European Union's Horizon 2020 research and innovation programme (Grant agreement No. ERC-2019-ADG 884902 SoftML).
\end{acknowledgments}

\section*{Conflict of Interest}
The authors have to conflicts to disclose.

\section*{Author Contribution}
\textbf{Thijs ter Rele}: Conceptualization (equal); Data Curation (lead); Formal Analysis (lead); Investigation (lead); Methodology (equal); Software (lead); Visualization (lead); Writing - original draft (lead); Writing - review \& editing (equal). 
\textbf{Ren\'{e} van Roij}: Conceptualization (equal);  Investigation (supporting); Methodology (equal); Visualization (supporting);   Writing - review \& editing (equal). 
\textbf{Marjolein Dijkstra}: Conceptualization (equal), Funding acquisition (lead); Investigation (supporting); Methodology (equal); Supervision (lead);  Visualization (supporting);  Writing - review \& editing (equal). 

\section*{Data Availibility Statement}
The data that support the findings of this study are available from the corresponding author upon reasonable request.

\bibliography{bibfile}

\clearpage
\appendix
\section{Molecular Dynamics Simulation Overview}
\noindent All simulations presented in this paper are molecular dynamics (MD) simulations performed using the LAMMPS software package.\cite{LAMMPS} The simulations are executed in a box of volume $V$ with periodic boundary conditions. During the simulation set-up, hard-sphere overlaps between particles are prevented through an energy-minimization step using a steepest descent algorithm. The MD simulations are performed in the $NVT$ ensemble, where the equations of motion are integrated using a Verlet 
scheme. A Nos\'{e}-Hoover thermostat is employed to maintain 
an average kinetic energy per particle of $\langle E_{kin} \rangle = 3 k_B T/2$.
All particles in the simulations have mass $m$; this includes both colloids and ions. The diameter of the ions is twenty times smaller than the colloid diameter, i.e.  $\sigma_i/\sigma = 0.05$. The characteristic MD time unit is  $\tau_{MD} = \sqrt{m \sigma^2/(k_B T)}$.

\section{Symmetry Functions and Gradients}
The constructed ML potentials are expressed in terms of weighted symmetry functions that characterize  the local environment of each particle. These symmetry functions, originally introduced by Behler and Parrinello,\cite{Behler2007}  come in two forms: the two-body symmetry functions $ G^{(2)}(i)$, which depend only on the radial distance $R_{ij} = |\mathbf{R}_j -\mathbf{R}_i|$ between particle $i$ and its neighbors, and the three-body symmetry functions $ G^{(3)}(i)$, which also depends on the angular arrangement of particle $i$ relative to two other particles  $j$ and $k$. 
The two-body symmetry function is defined as  
\begin{align}\label{G2}
    G^{(2)}(i) &= \sum_j e^{-\gamma(R_{ij} - R_s)^2} f_c(R_{ij}), 
\end{align}
where the sum runs over all neighboring particles $j$, and where $\gamma$ and $R_s$ are optimization parameters (see below) that determine the width and center of the Gaussian in $R_{ij}$, respectively. This symmetry function includes a cut-off function $f_c(R_{ij})$, defined as   
\begin{equation}
    f_c(R_{ij}) =  \begin{cases}
\tanh^3{\left(1 - R_{ij}/R_c\right)} & \text{for } R_{ij} \leq R_c; \\
0 &\text{for } R_{ij} > R_c,
\end{cases}
\end{equation}
which smoothly decays to zero as $R_{ij}$ approaches the  cut-off distance $R_c$. The three-body symmetry function is given by 
\begin{align}\label{G3}
    G^{(3)}(i) = & 2^{1-\xi} \sum_{j, k\neq i} \left(1 + \lambda \cos{\theta_{ijk}} \right)^{\xi} \nonumber
    \\ & e^{-\gamma \left(R_{ij}^2 + R_{ik}^2 + R_{jk}^2\right)} f_c(R_{ij}) f_c(R_{jk}) f_c(R_{ik}),
\end{align} where the sum runs over all distinct pairs 
$j,k\neq i$ within the cut-off distance $R_c$ from particle $i$, and where $\theta_{ijk}$ is the angle between the vectors ${\bf R}_{ij}={\bf R}_j-{\bf R}_i$ and ${\bf R}_{ik}={\bf R}_k-{\bf R}_i$. The optimization parameters $\gamma$, 
and $\lambda$ control the radial and angular resolution of the symmetry functions, respectively, while the parameter $\xi$ only takes  values of $+1$ or $-1$.

In this work, each ML potential is constructed from a set of 20 symmetry functions (SFs), selected to optimally force-match the ML potential to the training data. These 20 SFs are selected from a larger pool of 161 candidate symmetry functions, each characterized by seven optimization parameters

\noindent For the two-body symmetry functions $G^{(2)}$ the optimization parameters are: 
$\gamma \sigma^{2}\in \{0.01, 0.1, 1, 2, 4, 8, 16\} $; 
$R_s/\sigma \in \{0.0, 0.1, 0.2, 0.3, 0.4, 0.5, 0.6, 0.7, 0.8, 0.9, 1.0\} \sigma$.  For the three-body symmetry functions $G^{(3)}$ the parameters are: 
$\gamma \sigma^2 \in \{0.01, 0.1, 1, 2, 4, 8, 16\}$; 
$\lambda \in \{1, 2, 4, 8, 16, 32\}$; $\xi \in \{1, -1\}$. 
Throughout this work we use a cut-off radius of $R_c = 2.5 \sigma$.
For additional details on the construction of the ML potentials and the  symmetry-function framework, we refer the reader to Ref. \citenum{terRele2025}.

\section{Grand Canonical Monte Carlo simulations}\label{sec:sup:gcmc}
In Section \ref{sec:electrolyte_salt} we perform simulations in which the salt is treated grand-canonically. In this approach, salt is added to and removed from the simulation box as counterion-coion pairs using a Grand Canonical Monte Carlo (GCMC) scheme in the MD simulations.\cite{FrenkelenSmit} The ions are inserted in charge-neutral pairs of counter- and coions.

The acceptance rate of this move depends on two factors. The first  is the difference in potential energy $\mathcal{U}$ of adding or removing an ion pair. If the energy cost of an insertion or deletion is small (or even negative), the move is more likely to be accepted. The second factor is the chemical potential $\mu$ of the salt pair.  Within the implemented GCMC framework, the insertion or removal of a charge-neutral ion pair  is accepted with probabilities
\begin{align}
\text{acc}(N_{-};N_{+},N\rightarrow N_{-}+1;N_+ +1;N) & =  \nonumber \\
& \hspace{-5cm} \text{min}\Big[1,
\frac{V }{\Lambda^3 (N_{+}+1\big) } \frac{V }{\Lambda^3 (N_{-}+1\big) } \exp{[\beta\mu]}\times \nonumber \\
& \hspace{-5cm}\exp{\big[-\beta(\mathcal{U}(N_{+}+1;N_-+1;N) - \mathcal{U}(N_{+};N_-;N) }) \Big]; \\
\text{acc}(N_{-};N_{+},N \rightarrow N_{-}-1;N_{+}-1,N) & = \nonumber \\
&\hspace{-5cm}\text{min}\Big[1, 
\frac{\Lambda^3 N_{-}}{V} \frac{\Lambda^3 N_{+}}{V} \exp{[-\beta\mu]}\times \nonumber \\
& \hspace{-5cm} \exp{\big[-\beta( \mathcal{U}(N_{+}-1;N_--1;N) -\mathcal{U}(N_{+};N_-;N))\big]} \Big], 
\end{align}
where $\mathcal{U}(N_{+};N_-;N)$ represents the potential energy associated with a configuration of $N_{-}$ counterions, $N_+$ coions, and $N$ colloids.\cite{FrenkelenSmit}
The thermal wavelength $\Lambda$ is chosen to be equal to the colloid diameter $\Lambda = \sigma$. 

In this work, we use the inverse screening length $\kappa = \sqrt{4 \pi \lambda_B c_s} $ as a measure of the screening strength in the system, where $c_s = (N_+ + N_-)/V$ denotes the reservoir salt concentration, with $N_+$ and $N_-$ the (equal) number of counter- and coions in a reservoir volume $V$, respectively. To  determine the reservoir concentration, we  perform a simulation without colloids, i.e. at $\eta = 0$, at certain values of $\mu$ and $\sigma/\lambda_B$, allow the system to  equilibrate, and then determine the average number of ions in the simulation box. From this average, we obtain the concentration $c_s$ and the inverse screening length $\kappa$. 

\section{State diagrams}
\begin{figure}[b]
\includegraphics[width=0.95\linewidth]{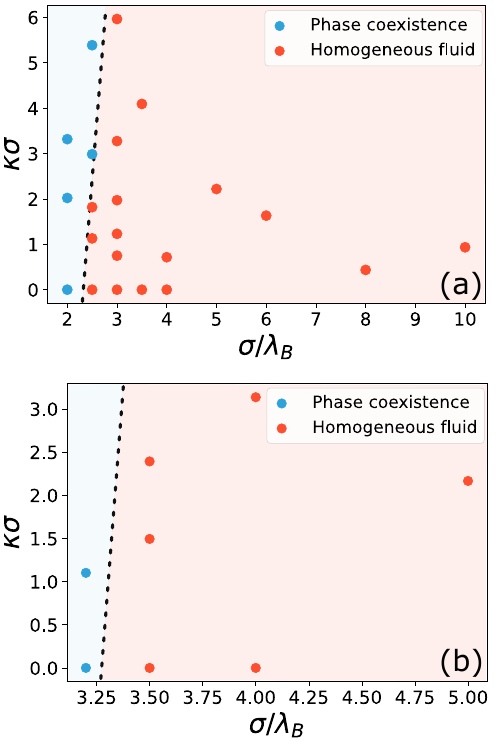}
\caption{State diagram in the temperature $\sigma/\lambda_B$ - screening parameter $\kappa\sigma$ representation, indicating the regime in which gas-liquid or gas-solid phase coexistence occurs, as well as the regime in which homogeneous fluid states are found, for colloids with valency (a) $Z = 100$ and (b) $Z =200$.
{The dashed line is a guide to the eye and should not be taken as the precise boundary between the two phases.}}
\label{fig:salt_coex_over_Z100_Z200_large}
\end{figure}

Fig. \ref{fig:salt_coex_over_Z100} presents the state diagram in the temperature $\sigma/\lambda_B$ - screening parameter $\kappa\sigma$ representation, indicating the regime in which gas-liquid or gas-solid phase coexistence occurs as well as  where homogeneous fluid states are found for colloids with a charge valency of  $Z =100$. In Fig. \ref{fig:salt_coex_over_Z100_Z200_large},  we show this state diagram over an extended   temperature range $\sigma/\lambda_B$, in (a) for $Z =100$ and in (b) for $Z=200$. The steep slope of the melting line in the representation of Fig. \ref{fig:salt_coex_over_Z100_Z200_large}(a) and (b) shows that salt-induced melting of a crystal or salt-induced crystallization of a dilute colloidal gas phase is difficult to achieve away from the salt-free melting line $\sigma/\lambda_B=(Z/13.7)^{0.5}$.

\section{ML-potential Experiment}

To directly test the PB-based theory of Zoetekouw et al. on like-charged spinodal instabilities,\cite{Zoetekouw2006prl} we trained an ML potential for a parameter set investigated experimentally by Monovoukas and Gast,  at which like-charged phase separation was reported.\cite{Monovoukas1989} For the same parameter set, large regions of phase coexistence were predicted by PB-based theories.\cite{Russ2002, Zoetekouw2006}
In particular, Zoetekouw et al.  constructed a phase diagram for these parameters using a PB-based framework that incorporates so-called ``volume-terms''. These volume-term contributions can be interpreted as an effective representation of the  many-body character of the colloidal interactions, projected onto a single density-dependent potential-energy  contribution. The attractive potential contribution arising from these volume terms leads to a  broad coexistence  region in the PB-based phase diagram, extending  from low to high densities.

\begin{figure}[ht]
    \includegraphics[width=0.9\linewidth]{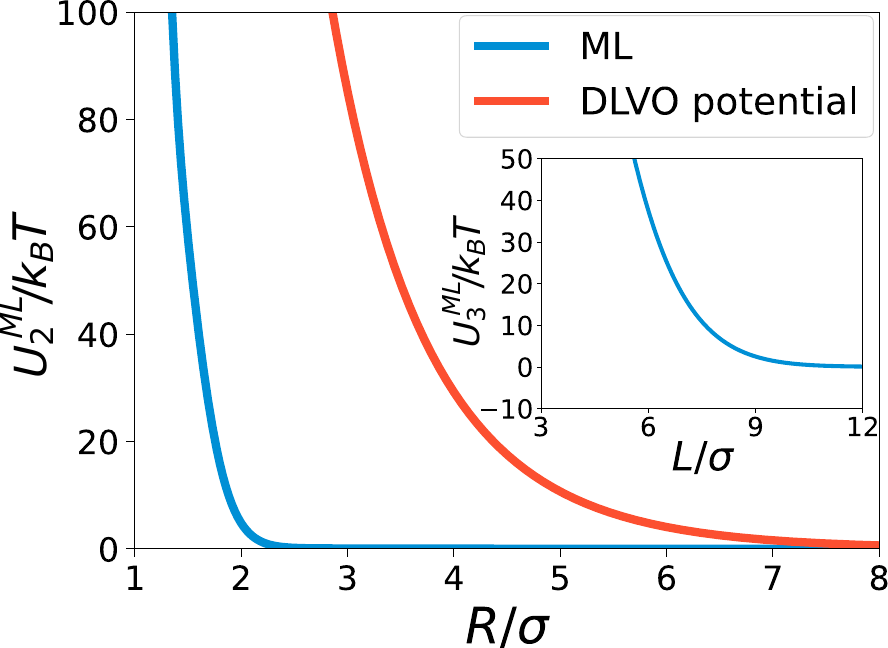}%
 \caption{ ML potential $U^{ML}(R)$ for charged colloids with valency $Z=1200$, at effective temperature $\sigma/\lambda_B = 185.3$ and screening parameter $\kappa \sigma =0.783$, with ion-to-colloid size ratio $\sigma_i/\sigma = 0.05$. The ML model was trained on 240 configurations of 14 colloids at packing fractions in the range $\eta \in \lbrack0.001, 0.45\rbrack$. The two-body potential component $U_2^{ML}(R)$ is plotted in the main figure, and is compared to an equivalent DLVO potential for renormalized charge $Z^* = 663$, determined using the  charge renormalization procedure of Alexander et al.\cite{Alexander1984} The three-body component $U_3^{ML}(R)$ of the potential for three particles arranged at an equilateral distance is shown in the inset, where $L = R_{12} + R_{23} + R_{13}$ denotes the total perimeter of the triangle connecting the particles.
 }
  \label{fig:app:potential-Monovoukas-Gast}
\end{figure}

We expect the ML framework to capture the attractive many-body character of the colloidal  interactions via an attractive three-body contribution. In order to further compare the experimental results with  the PB-based theory, we trained an ML potential at the same parameter set.  The resulting potentials are  presented in Fig. \ref{fig:app:potential-Monovoukas-Gast}, where we plot the two-body component of the ML potential $U^{2}_{ML}$  in the main figure and the three-body component  $U^{3}_{ML}$ in the inset. 
The resulting interaction is purely repulsive, as  both the two-body and three-body components exhibit  only positive  contributions at all colloid separations. The repulsive range of the two-body potential is significantly  shorter than that of the corresponding DLVO potential,  this is compensated  by the relatively long-ranged three-body repulsions in the ML description. 
Simulations performed with this potential, starting either from a homogeneous fluid or from a crystalline configuration, consistently relax to a homogeneous final state. There is no attractive potential-energy contribution present that could  induce a spinodal instability. This observation is consistent with the results presented in  Section \ref{sec:asymmetric-electrolyte}, where no phase separation was found at these relatively high effective temperatures.

\end{document}